\documentclass{moriond}

\def\be{\begin{equation}}
\def\ee{\end{equation}}
\def\bea{\begin{eqnarray}}
\def\eea{\end{eqnarray}}

\newcommand{\Photo}{\includegraphics[height=35mm]{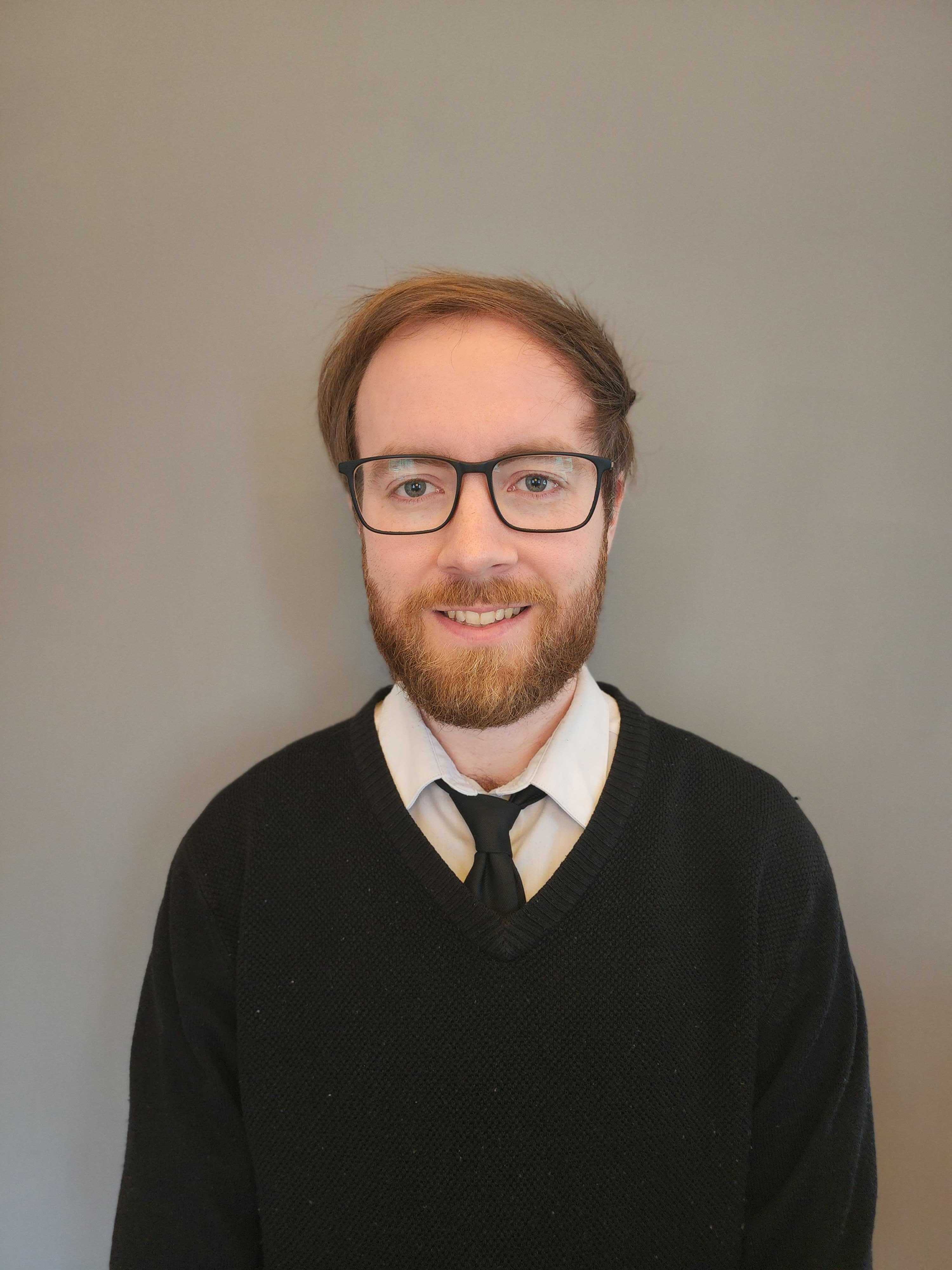}}

\begin{document}
\vspace*{4cm}
\title{CMB Spectral Distortions: A Multimessenger Probe of the Primordial Universe}

\author{Bryce Cyr}

\address{Jodrell Bank Centre for Astrophysics, School of Physics and Astronomy, The University of Manchester, Manchester M13 9PL, UK}

\maketitle\abstracts{The frequency spectrum of the cosmic microwave background (CMB) is a relatively untapped source of data which can allow us to peer beyond the surface of last scattering. Small deviations away from a perfect blackbody shape will encode valuable information about the state of the primordial Universe which may not be accessible by other means. Here, we briefly review some key science goals of CMB spectral distortions, with an emphasis on how future generations of experiments can be used in tandem with complementary observational probes to perform model discrimination of exotic physics scenarios. We focus here on synergies between spectral distortions, gravitational waves, and $21$cm cosmology.}

\section{CMB Spectral Distortions - A Brief Introduction}
In the mid 90s, the cosmic microwave background community was perched on the edge of their seats, anxiously awaiting results from the COBE experiment which would help to usher in an age of precision cosmology. The satellite contained three major instruments with lofty science goals: DIRBE \cite{Schlegel1997}, a precision dust mapper, DMR \cite{COBE1992}, searching for temperature anisotropies, and FIRAS \cite{Fixsen1996}, to provide a precise measurement of the frequency spectrum of background radiation. Following the COBE results, it was clear that a treasure trove of temperature anisotropies were lurking just below the DMR sensitivity, thus beginning a rich legacy of anisotropy studies which included the WMAP \cite{WMAP2012} and Planck \cite{Planck2018} satellites, whose results eventually culminated into the concordance ($\Lambda$CDM) model of cosmology that we have today.

For the past three decades, the community has mounted an impressive and exhaustive study of CMB temperature and polarization anisotropies. Current experimental efforts are keen to measure B-mode polarization on large angular scales, in hopes of detecting the imprints of primordial gravitational waves. Along the way, these measurements offer improved constraints on various models of inflation and other processes in the early Universe capable of sourcing tensor modes. If such a detection is made, a new observational targets will need to be considered. One prime candidate for future missions would be spectral distortions (SDs) of the CMB.

In the early Universe, rapid interactions between the photons and baryons efficiently thermalized the background, enforcing a blackbody shape to the photon spectrum. Thermalization, however, is not an instantaneous process. In an expanding Universe, the microphysical interactions necessary to maintain thermal equilibrium (typically Bremsstrahlung, Compton, and double Compton scattering) damp over time, eventually freezing out at redshifts long before the formation of the CMB. Non-thermal energy release after freeze out is capable of leaving a characteristic imprint on the CMB frequency spectrum, known as a CMB spectral distortion. For this discussion, we will focus on global (or sky-averaged) distortions, but note that recent progress has also been made on anisotropic signatures \cite{SD-Ani-3}, where $\mu-T$ correlations provide another exciting new window into the early Universe.

At redshifts of $z \simeq 10^6$, interactions which allow for the creation and destruction of photons freeze out, after which it becomes possible to source SDs. Non-thermal injections around this time are still capable of redistributing their energy efficiently through rapid Compton scattering events, which instead drives the equilibrium distribution of photons to a Bose-Einstein type with a chemical potential ($\mu$) proportional to the fractional energy release. This $\mu$-era lasts until $z\simeq 10^5$, when the efficacy of Compton scattering begins to waver. At this point, a $y$-type spectral distortion is produced which in principle can be sourced even post-recombination. For pre-recombination injections, one can estimate the amplitude of these $\mu$ and $y$ distortions using a Green's method approach \cite{Chluba2016}
%
\begin{eqnarray}
\mu &\simeq& 1.401 \int_0^{\infty} {\rm d}z \left(\frac{1}{\rho_{\gamma}} \frac{{\rm d}\rho_{\rm inj}}{{\rm d}z} - \frac{4}{3N_{\gamma}} \frac{{\rm d}N_{\rm inj}}{{\rm d}z}\right) \mathcal{W}_{\mu}  \approx  1.401\left.\left( \frac{\Delta \rho_{\gamma}}{\rho_{\gamma}} - \frac{4}{3} \frac{\Delta N_{\gamma}}{N_{\gamma}}\right)\right|_{\mu{\rm-era}},\\
y &\simeq&  \frac{1}{4} \int_0^{\infty} {\rm d}z \frac{1}{\rho_{\gamma}} \frac{{\rm d}\rho_{\rm inj}}{{\rm d}z} \, \mathcal{W}_{y}  \approx  \frac{1}{4} \left. \frac{\Delta \rho_{\gamma}}{\rho_{\gamma}} \right|_{{\rm y-era}}.
\end{eqnarray}
%
Here, $\rho_{\gamma}$ and $N_{\gamma}$ are the CMB energy and number densities respectively, while $\mathcal{W}_{\mu/y}$ pick out the redshift regimes where the different distortions are active \cite{Cyr2023}. In principle, for $10^4 \leq z \leq 10^5$ residual non $\mu/y$ distortions can be sourced which allow one to reconstruct the time dependence of the injection process. For more detailed calculations, the \texttt{CosmoTherm} \cite{Chluba2011} codebase allows one to study both standard and exotic heating scenarios in exquisite detail.

COBE/FIRAS placed upper bounds on the distortion parameters of \footnote{A recent re-analysis of the FIRAS data has resulted in a factor of $\simeq 2$ improvement on $\mu$ \cite{Bianchini2022}.} $|\mu| \leq 9 \times 10^{-5}$  and $y \leq 1.5 \times 10^{-5}$. While a dedicated experimental follow-up to this measurement has not yet materialized, the landscape is beginning to heat up. In the coming years, data from the TMS experiment \cite{Jose2020TMS} will provide new limits on the spectrum between the ARCADE-2 \cite{Fixsen2009} and FIRAS bands, forecasted to reach a nominal $y$-distortion limit of $y \leq 10^{-6}$. Additionally, the balloon-bourne experiment BISOU \cite{BISOU} is expected to be given the green light, planning to target the more traditional CMB band with more than an order of magnitude improvement in sensitivity. More futuristically, the Voyage2050 programme is a call-to-action by the European Space Agency (ESA) for L-class mission proposals, where spectral distortions have been named as a potential science target \cite{Chluba2021Voyage}. In preparation for the upcoming call for proposals, the community is making strides in understanding the foreground challenges and opportunities \cite{Abitbol2017}, as well as reviewing experimental designs, with the expectation that a sensitivity of $|\mu| \leq 2 \times 10^{-8}$ can be reached using a PIXIE-type design \cite{Kogut2024}, an enormous improvement over current limits.

So what science targets exist for spectral distortions? Within the vanilla $\Lambda$CDM setup, one expects global $y$ distortion of $\simeq 10^{-6}$ coming from both heating at reionization, as well as the sky-average of hot galaxy clusters which exhibit the Sunyaev-Zeldovich effect. This post-recombination signal should be in reach of BISOU (and perhaps even TMS), and would stand as the first ever detection of a global CMB spectral distortion.

\begin{figure*}
\centering 
\includegraphics[width=1.0\columnwidth]{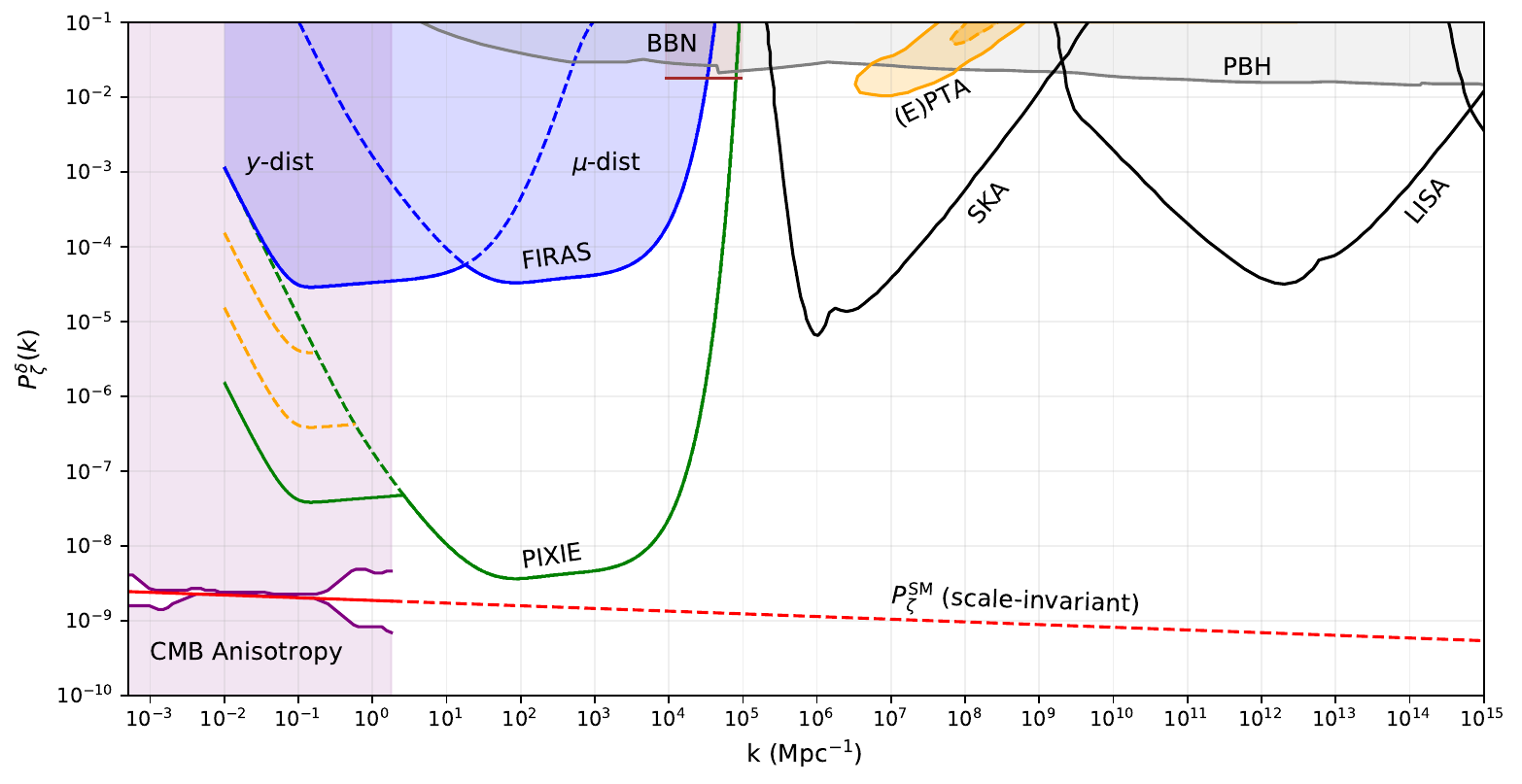}
\caption[]{Constraints (filled contours) and forecasts for $\delta$-function enhancements to the primordial scalar power spectrum. Note that spectral distortions are an integrated constraint, and that integration over the $\mu$ window of $P_{\zeta}^{\rm SM}$ yields a distortion which should be detectable by a PIXIE-type instrument. Figure adapted from Cyr \textit{et al.}\cite{Cyr2023}.}
\label{fig:SD-PPS}
\end{figure*}

For primordial $\mu$ distortions, the largest expected signal comes from the Silk damping of primordial density fluctuations after they re-enter the cosmological horizon. Upon re-entry, these modes oscillate until they reach a characteristic damping scale, at which point they diffuse and dump their energy into the background plasma. This process, which can be viewed as the mixing of patches with different temperatures \footnote{The sum of two blackbody spectra with different $T$ is in fact not a blackbody itself.}, leads to a characteristic distortion \cite{Chluba2012} \cite{Cyr2023} with amplitude $\mu \simeq 10^{-8}$, within reach of a next generation space mission. Modes which deposit their energy during the $\mu$ window lay in the (comoving) range $50 \, {\rm Mpc}^{-1} \leq k \leq 10^4 \, {\rm Mpc}^{-1}$, which means that SDs are sensitive to the primordial scalar power spectrum on smaller scales than those probed by CMB anisotropy missions such as Planck. The predicted amplitude of $\mu$ is determined by assuming the spectrum remains roughly scale-invariant, so even non-detection by a PIXIE-type instrument could provide us with fundamental insights into the inflationary potential or other conditions present in the very early Universe. As illustrated in Fig.~\ref{fig:SD-PPS}, SDs offer the most sensitive probe to density fluctuations on these small scales. 
%
%

In addition to these $\Lambda$CDM signatures \cite{Chluba2016}, data from a next generation experiment would allow for an unprecedented jump in constraining power on many exotic physics models, such as primordial black holes \cite{Carr2020}, cosmic strings \cite{Cyr2023-SCS-Constraints}, decaying/annihilating dark matter \cite{Bolliet2020}, and primordial magnetic fields \cite{Vachaspati2020}. The theory space of models which induce spectral distortions at a level detectable by e.g. PIXIE is vast, and many such models also produce signatures across other observational windows. As data collection continues across this array of cosmological experiments, it becomes increasingly important to leverage new information in order to perform robust model discrimination of new physics scenarios. CMB spectral distortions can play a key role in helping to disentangle the various properties of astrophysical and cosmological anomalies which show up in the data, including the presence of stochastic gravitational wave backgrounds \cite{Cyr2023}, and additional low-frequency photon backgrounds \cite{Cyr-SPH-Analytic} such as the one detected by ARCADE-2 \cite{Fixsen2009} and LWA \cite{DT2018}.

\section{A New CMB Probe of Gravitational Waves}
The detection of gravitational waves (GWs), both in the form of a stochastic background (as discovered recently by the pulsar timing array consortium \cite{NANOGravDetection2023}), as well as transient bursts (courtesy of the LIGO/Virgo collaboration \cite{LIGOScientific2016}), has opened a new observational window into the Universe. While we have a robust understanding of the black hole/neutron star progenitors responsible for GW bursts, things are less clear for the stochastic background. The spectrum of this background has been reported to be blue tilted with spectral index $\gamma \simeq 3.2$, roughly $3\sigma$ away from the expected theory prediction for a population of inspiraling supermassive black hole binaries (SMBHB).

\begin{figure*}
\centering 
\includegraphics[width=1.0\columnwidth]{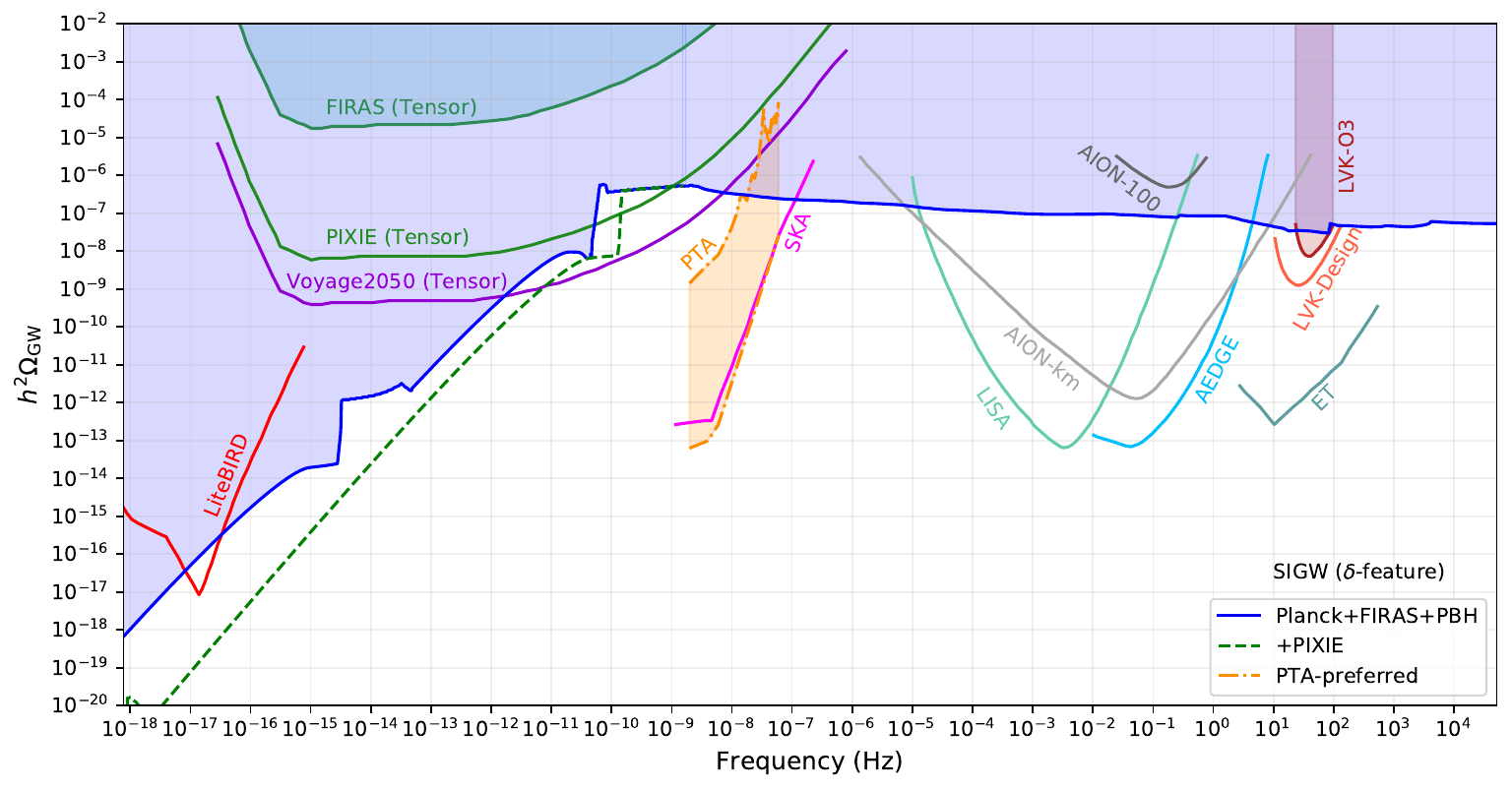}
\caption[]{An overview of gravitational wave constraints from a variety of instruments. The shaded blue represents the exclusions region for $\delta$ function enhancements to the scalar power spectrum (with the dashed green indicating a PIXIE forecast). Constraints marked (Tensor) come from direct dissipation \cite{Kite2020} of these modes during the $\mu$ era, leading to model independent constraints that ``bridge the gap" between B-mode searches and direct detection.}
\label{fig:SD-GW}
\end{figure*}

This discrepancy has motivated many \footnote{Including the NANOGrav collaboration \cite{NANOGrav2023Exotic} themselves!} to take a closer look at creative explanations, with exotic models such as domain walls, (some) cosmic string models, and scalar-induced gravitational waves (SIGWs) appearing to be favoured over the SMBHB explanation. While a full analysis of each exotic model is perhaps unreasonable, a more straightforward question one can ask is if this gravitational wave background was sourced primordially (before recombination), or more locally (post-recombination). The detection of B-mode polarization is one way to determine this, but is unfortunately only possible if the spectrum of gravitational waves extends to ultra-low frequencies (see the ``LiteBIRD" forecast in Fig.~\ref{fig:SD-GW}). Spectral distortions, on the other hand are capable of probing a much wider region of GW parameter space through both model dependent and independent effects.

As a case study, lets consider the scenario in which a $\delta$-function enhancement \footnote{While unphysical, this setup provides the qualitative picture while avoiding unnecessary numerical details.} to the primordial scalar power spectrum is present at some small scale $k_{\rm *, S}$ with amplitude $A_{\zeta}$. At second order in general relativity, it is well known that these scalar perturbations will interact, acting as a source term for tensor modes which are then free to propagate \cite{Ananda2006}. The cost for this is that the amplitude of gravitational waves is suppressed, $\Omega_{\rm GW} \propto A_{\zeta}^2$, implying that one needs rather large scalar enhancements to produce a sizeable background. The resultant GW spectrum can be solved analytically in the case of $\delta$ enhancements, consisting of a resonant peak at $k_{\rm *, GW} =(2/\sqrt{3}) k_{\rm *, S}$ during radiation domination and a decaying power law in the IR. 

As we witnessed in Fig.~\ref{fig:SD-PPS}, scalar enhancements at $k \leq 10^4 \, {\rm Mpc}^{-1}$ can be constrained quite tightly through CMB anisotropy and spectral distortion measurements. By saturating the bounds from these measurements, as well as the PBH curve in Fig.~\ref{fig:SD-PPS}, we can map these constraints from the scalar power spectrum onto the gravitational wave parameter space, yielding the shaded blue region in Fig.~\ref{fig:SD-GW}. This mapped constraint tells us that if we were to detect gravitational waves within the shaded region, we can rule out $\delta$-induced gravitational waves as a possible source. The green dashed contour shows how these constraints would improve with a PIXIE-type experiment. The mapping of these constraints from more realistic scalar power spectrum enhancements represents an important future avenue of study. Generally speaking, CMB spectral distortions tend to be more constraining with wider scalar enhancements \cite{Cyr2023}, so we expect this method to be more restrictive for physical scenarios.

Spectral distortions are also capable of constraining the presence of tensor modes in a model independent way, provided that they are sourced at high enough redshift ($z_{\rm src} \geq 10^6$). The presence of gravitational waves generates polarization fluctuations in the CMB. This in turn mediates the mixing of different patches of the Universe through free-streaming effects, producing a spectral distortion in a similar way to the Silk damping effect described above \cite{Kite2020}. In contrast to Silk damping, the amplitude of the distortion is typically much weaker. This is partially offset by the fact that distortions can be generated by a wider range of gravitational wavenumbers ($1 \, {\rm Mpc}^{-1} \leq k_{\rm GW} \leq 10^6 \, {\rm Mpc}^{-1}$) due to the fact that unlike scalar perturbations which decay rapidly once crossing the damping scale, tensor modes persist through to the present day. Model independent constraints on primordial gravitational waves are marked by the ``Tensor" label in Fig.~\ref{fig:SD-GW}.

\section{Soft Photon Heating and 21cm Cosmology}
A little over a decade ago, the ARCADE-2 experiment \cite{Fixsen2009} claimed the detection of an anomalous radio background present in the Rayleigh-Jeans tail of the CMB. A few years later, this low frequency spectral distortion was also discovered in data taken by the LWA \cite{DT2018}, helping to establish this radio background as a \textit{bona fide} anomaly, whose origin remains a topic of much debate \cite{RSBworkshop2022}. The brightness temperature of the radio background exhibits a power law departure from the CMB at roughly $\nu \leq 1$ GHz, and has been parameterized as \cite{Cyr2023-RSB}
%
\begin{equation}
    T_{\rm RB}(\nu) \simeq 1.230 \, {\rm K} \left( \frac{\nu}{{\rm GHz}}\right)^{-(\gamma -1)},
\end{equation}
%
where $\gamma = 3.555$ is the spectral index of the intensity spectrum. While the progenitor of such a background remains a mystery, a familiar question can be asked: was the radio background sourced in the dark ages ($30 \leq z \leq z_{\rm rec}$\footnote{Radio photons produced before recombination are quickly absorbed and reprocessed by the primordial plasma, so it is in general not possible to source this background at $z \geq z_{\rm rec}$.}), or at later times ($z \leq 30$)? A common misconception is that low frequency photon backgrounds have a negligible impact on the evolution of cosmological quantities such as the matter temperature. When sourced, radio backgrounds are expected to extend down to the plasma frequency. For sufficiently steep spectral indices, a large fraction of energy will reside at ultra-low frequencies, where they are susceptible to efficient free-free absorption (inverse Bremsstrahlung) effects. This can act as an impressive source of energy injection into the hydrogen gas during the dark ages, leading to a wide range of observable effects \cite{Cyr-SPH-Analytic}. 

\begin{figure*}
\centering 
\includegraphics[width=0.48\columnwidth]{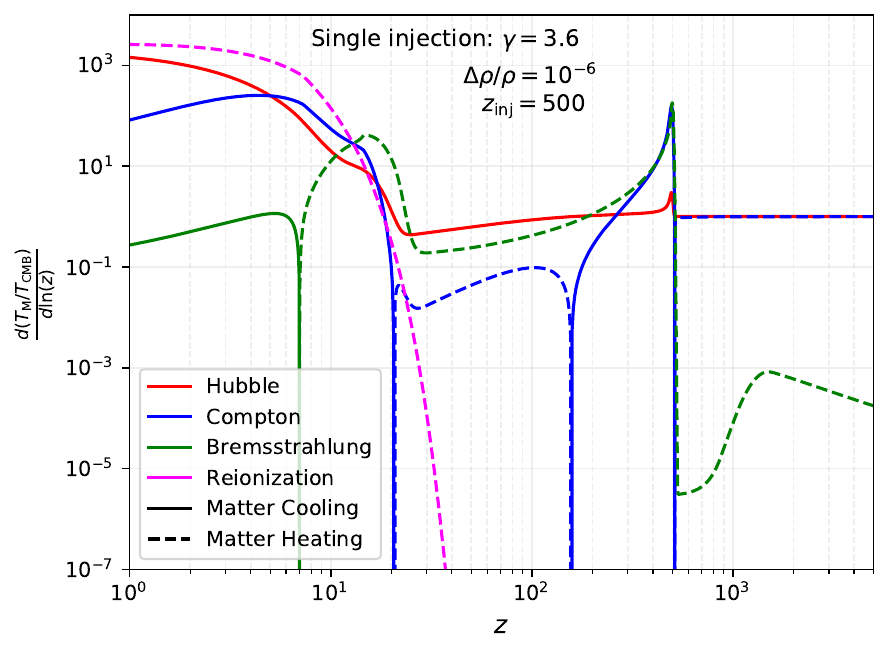}
\hspace{4mm}
\includegraphics[width=0.48\columnwidth]{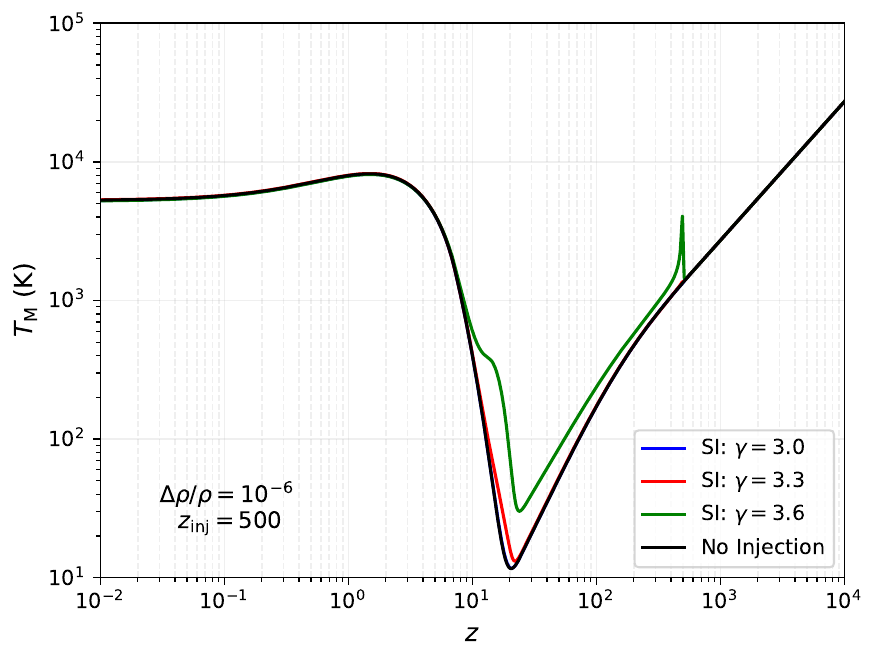}\\
\includegraphics[width=0.48\columnwidth]{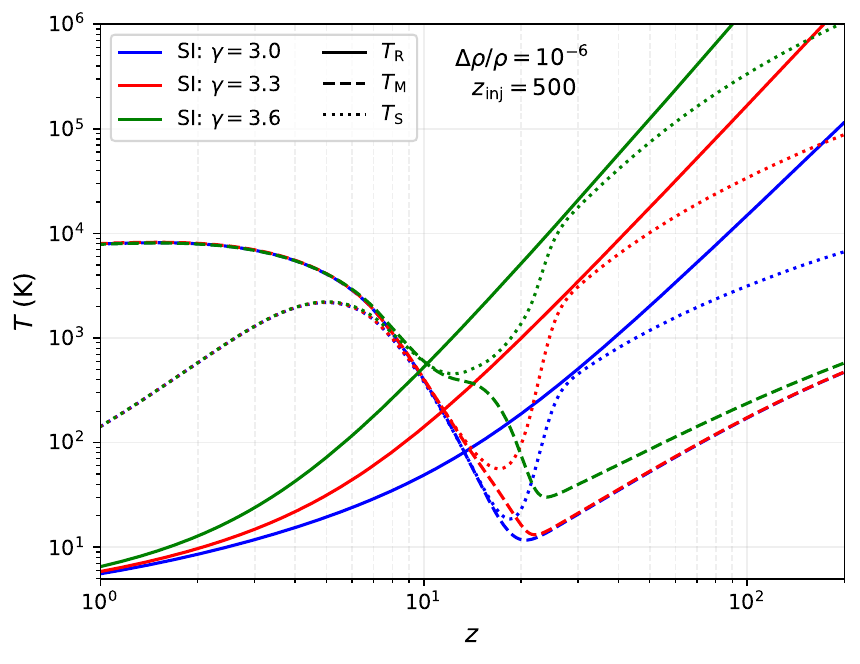}
\hspace{4mm}
\includegraphics[width=0.48\columnwidth]{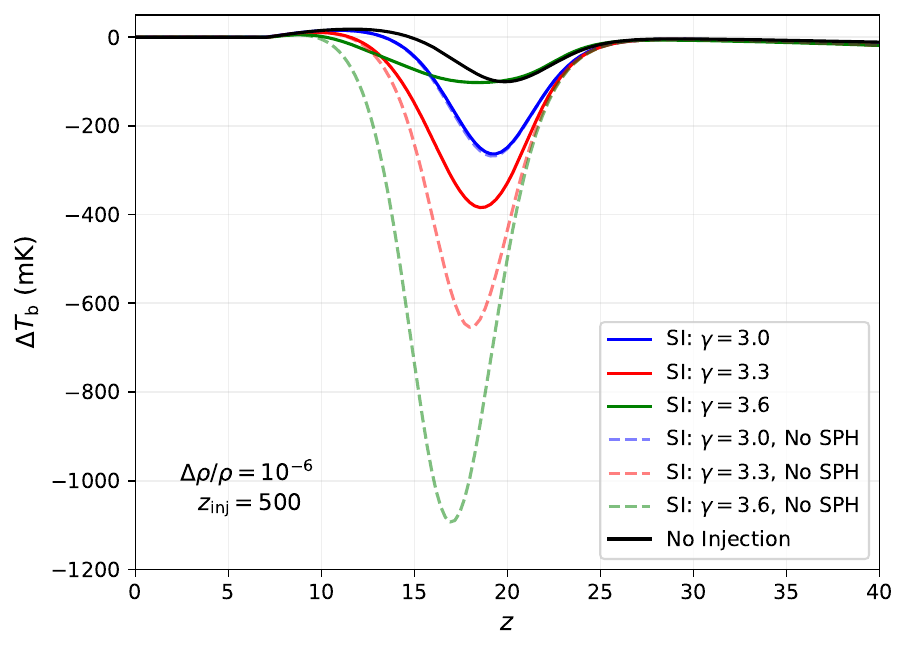}
\caption[]{Upper Left: Relative contributions of each of the terms in Eq.~(\ref{Eq:heating-rate}) to the global matter temperature for a fiducial radio background injection with a synchrotron-like spectrum at $z_{\rm inj} = 500$. Free-free (Bremsstrahlung) heating becomes dominant around cosmic dawn. Upper Right: Evolution of the matter temperature with background injections when varying $\gamma$. Lower Left: The evolution of the radiation, matter, and spin temperatures in the presence of a quasi-instantaneous radio injection at $z_{\rm inj} = 500$. Lower Right: The global differential brightness temperature for different spectral indices with (solid) and without (dashed) soft photon heating included.}
\label{fig:zinj500-Tm}
\end{figure*}

To highlight this ``soft photon heating" effect, work was carried out which considered both quasi-instantaneous and continuous injections of a radio background at high redshifts, studying the subsequent evolution of various cosmological quantities such as the matter temperature ($T_{\rm M}$) and ionization fraction ($X_{\rm e}$). These radio background were parameterized by an amplitude ($\Delta \rho/\rho$), as well as a spectral index ($\gamma$). The evolution of the matter temperature follows
%
\begin{equation} \label{Eq:heating-rate}
    \frac{{\rm d} T_{\rm M}}{{\rm d} z} = \frac{2 T_{\rm M}}{1+z} + \frac{X_{\rm e}}{1+ X_{\rm e} + f_{\rm He}} \frac{8 \sigma_{\rm T} \rho_{\gamma}}{3 m_{\rm e} c} \frac{T_{\rm M} - T_{\rm CMB}}{H(z) (1+z)} + \frac{{\rm d} T_{\rm ff}}{{\rm d} z},
\end{equation}
%
where the first term comes from the adiabatic expansion of the gas, the second from Compton cooling effects, and the third from free-free absorption (soft photon heating). The upper left panel of Fig.~\ref{fig:zinj500-Tm} showcases the relative contribution of each of these terms for a quasi-instantaneous injection at $z = 500$, where it is clear that soft photon heating is important both near the injection time and at $z \approx 20$ due to an increase in the ionization fraction as the first stars begin to turn on. The right panel shows the full evolution of $T_{\rm M}$, where it is clear that for a synchrotron-type spectral index ($\gamma \simeq 3.6$), the gas will get significantly warmer at all redshifts between the time of injection and the conclusion of reionization.
%
%

A natural application of this soft photon heating effect is on the quantities of interest in $21$cm cosmology, such as the differential brightness temperature $\Delta T_{\rm b}$. This quantity acts as a measure of the contrast between the radiation temperature at $21$cm wavelengths ($T_{\rm R}$), and the so-called spin temperature ($T_{\rm S}$), through $\Delta T_{\rm b} \propto (1-T_{\rm R}/T_{\rm S})$. $21$cm photons are produced through a ``forbidden" spin-flip transition exhibited by neutral hydrogen. The intensity of these photons therefore depends on the relative occupation of the spin triplet state to the singlet, which is precisely what the spin temperature quantifies. $T_{\rm S}$ is sensitive to any process capable of populating the triplet state, such as through resonant absorptions of $21$cm photons or a hotter gas temperature. As seen in the upper panels of Fig.~\ref{fig:zinj500-Tm}, the presence of high redshift radio backgrounds will play a role in the evolution of these quantities, and these effects cannot be neglected when performing a computation of the spin temperature. 

In the lower left panel of Fig.~\ref{fig:zinj500-Tm}, the evolution of the radiation, matter, and spin temperatures are plotted with the same radio background injection considered above. Previous studies had made the incorrect assumption that the presence of extra radio backgrounds would produce only a negligible effect on the spin temperature evolution. We find that the main effect of soft photon heating is to reduce the contrast between the radiation and spin temperatures, which causes a significant suppression in the amplitude of $\Delta T_{\rm b}$. This is exemplified in the lower right panel of Fig.~\ref{fig:zinj500-Tm} where we compute $\Delta T_{\rm b}$ for a variety of spectral indices both with and without including soft photon heating. The comparison is rather dramatic, with the conclusion being that the presence of sufficiently steep radio backgrounds at cosmic dawn can cause the global $21$cm signal to be even more difficult to detect, contradicting previous claims in the literature. The impact of soft photon heating on the $21$cm power spectrum is currently an active area of research, with much phenomenology still waiting to be uncovered.

\section{Concluding Thoughts}
The cosmic microwave background has provided us with a wealth of knowledge thus far, mainly through precise measurements of the temperature and polarization anisotropies over a wide range of scales. The frequency spectrum, however, remains a relatively untapped resource that could serve as a major science target in the coming decades. 

While CMB spectral distortions are certainly worthy of study and observation in their own right, they are also able serve as a powerful tool to help discriminate between different models of the early and late time Universe, which can be notoriously difficult to test otherwise. Here we have highlighted just two examples of how SDs can be used to disentangle early and late time sources of gravitational waves, as well as extra radio backgrounds through their adverse effects on $21$cm cosmology. Much exciting work remains to be done, both in understanding the phenomenological implications of these ideas, and in elucidating further synergies.

\section*{Acknowledgments}

B.C. would like to acknowledge contributions from collaborators Jens Chluba and Sandeep Acharya. B.C. is supported by both an NSERC postdoctoral fellowship and the ERC Consolidator Grant CMB-SPEC (No. 725456). He would also like to thank the organizers of the 2024 Moriond Cosmology conference for the invitation to give the talk on which this article is
based.

\section*{References}

\bibliography{Lit}

\end{document}